\input harvmac
%\draftmode
%\def\IR{\relax{\rm I\kern-.18em R}}
%\input epsfDavid

\let\includefigures=\iftrue
\let\useblackboard=\iftrue
\newfam\black

%Figure Stuff
\includefigures
\message{If you do not have epsf.tex (to include figures),}
\message{change the option at the top of the tex file.}
\input epsf
\def\figin{\epsfcheck\figin}\def\figins{\epsfcheck\figins}
\def\epsfcheck{\ifx\epsfbox\UnDeFiNeD
\message{(NO epsf.tex, FIGURES WILL BE IGNORED)}
\gdef\figin##1{\vskip2in}\gdef\figins##1{\hskip.5in}% blank space instead
\else\message{(FIGURES WILL BE INCLUDED)}%
\gdef\figin##1{##1}\gdef\figins##1{##1}\fi}
\def\DefWarn#1{}
\def\figinsert{\goodbreak\midinsert}
\def\ifig#1#2#3{\DefWarn#1\xdef#1{fig.~\the\figno}
\writedef{#1\leftbracket fig.\noexpand~\the\figno}%
\figinsert\figin{\centerline{#3}}\medskip\centerline{\vbox{
\baselineskip12pt\advance\hsize by -1truein
\noindent\footnotefont{\bf Fig.~\the\figno:} #2}}
%\bigskip
\endinsert\global\advance\figno by1}
%%%
\else
\def\ifig#1#2#3{\xdef#1{fig.~\the\figno}
\writedef{#1\leftbracket fig.\noexpand~\the\figno}%
%\figinsert\figin{\centerline{#3}}\medskip
%\centerline{\vbox{\baselineskip12pt
%\advance\hsize by -1truein\noindent
%\footnotefont{\bf Fig.~\the\figno:} #2}}
%\bigskip\endinsert
\global\advance\figno by1} \fi

\def\journal#1&#2(#3){\unskip, \sl #1\ \bf #2 \rm(19#3) }
\def\andjournal#1&#2(#3){\sl #1~\bf #2 \rm (19#3) }

\def\ie{{\it i.e.}}
\def\eg{{\it e.g.}}

\noblackbox
%

% Something to deal with sub-sub-sections

\def\unlockat{\catcode`\@=11}
\def\lockat{\catcode`\@=12}

\unlockat
% Something to deal with sub-sub-sections

\def\newsec#1{\global\advance\secno by1\message{(\the\secno. #1)}
\global\subsecno=0\global\subsubsecno=0\eqnres@t\noindent
{\bf\the\secno. #1}
\writetoca{{\secsym} {#1}}\par\nobreak\medskip\nobreak}
\global\newcount\subsecno \global\subsecno=0
\def\subsec#1{\global\advance\subsecno
by1\message{(\secsym\the\subsecno. #1)}
\ifnum\lastpenalty>9000\else\bigbreak\fi\global\subsubsecno=0
\noindent{\it\secsym\the\subsecno. #1}
\writetoca{\string\quad {\secsym\the\subsecno.} {#1}}
\par\nobreak\medskip\nobreak}
\global\newcount\subsubsecno \global\subsubsecno=0
\def\subsubsec#1{\global\advance\subsubsecno by1
\message{(\secsym\the\subsecno.\the\subsubsecno. #1)}
\ifnum\lastpenalty>9000\else\bigbreak\fi
\noindent\quad{\secsym\the\subsecno.\the\subsubsecno.}{#1}
\writetoca{\string\qquad{\secsym\the\subsecno.\the\subsubsecno.}{#1}}
\par\nobreak\medskip\nobreak}

\def\subsubseclab#1{\DefWarn#1\xdef
#1{\noexpand\hyperref{}{subsubsection}%
{\secsym\the\subsecno.\the\subsubsecno}%
{\secsym\the\subsecno.\the\subsubsecno}}%
\writedef{#1\leftbracket#1}\wrlabeL{#1=#1}}% Macros for boxes
\lockat

\def\ie{{\it i.e.}}
\def\eg{{\it e.g.}}

%% MORE MACROS

\def\CN {{\cal N}}

\font\manual=manfnt \def\dbend{\lower3.5pt\hbox{\manual\char127}}

\def\IZ{\relax\ifmmode\mathchoice
{\hbox{\cmss Z\kern-.4em Z}}{\hbox{\cmss Z\kern-.4em Z}}
{\lower.9pt\hbox{\cmsss Z\kern-.4em Z}}
{\lower1.2pt\hbox{\cmsss Z\kern-.4em Z}}\else{\cmss Z\kern-.4em
Z}\fi}
\def\half{{1\over 2}}

\def\CN {{\cal N}}

% more macros, alphabetically

\def\IZ{\relax\ifmmode\mathchoice
{\hbox{\cmss Z\kern-.4em Z}}{\hbox{\cmss Z\kern-.4em Z}}
{\lower.9pt\hbox{\cmsss Z\kern-.4em Z}}
{\lower1.2pt\hbox{\cmsss Z\kern-.4em Z}}\else{\cmss Z\kern-.4em
Z}\fi}
\def\IB{\relax{\rm I\kern-.18em B}}
\def\IC{{\relax\hbox{$\inbar\kern-.3em{\rm C}$}}}
\def\ID{\relax{\rm I\kern-.18em D}}
\def\IE{\relax{\rm I\kern-.18em E}}
\def\IF{\relax{\rm I\kern-.18em F}}
\def\IG{\relax\hbox{$\inbar\kern-.3em{\rm G}$}}
\def\IGa{\relax\hbox{${\rm I}\kern-.18em\Gamma$}}
\def\IH{\relax{\rm I\kern-.18em H}}
\def\II{\relax{\rm I\kern-.18em I}}
\def\IK{\relax{\rm I\kern-.18em K}}
\def\IP{\relax{\rm I\kern-.18em P}}
\def\IQ{\relax\hbox{$\inbar\kern-.3em{\rm Q}$}}

\def\inbar{\,\vrule height1.5ex width.4pt depth0pt}

\font\cmss=cmss10 \font\cmsss=cmss10 at 7pt
\def\IR{\relax{\rm I\kern-.18em R}}

% Macros for boxes
%
%\def\boxit#1{\vbox{\hrule\hbox{\vrule\kern8pt
%\vbox{\hbox{\kern8pt}\hbox{\vbox{#1}}\hbox{\k
%\hbox{$\displaystyle #1$}\kern8pt}\kern8pt\vrule}\hrule}}}
%
%%% MACROS FOR BOX BOUNDARY CONDS
%%% FROM KAWAI ET AL

\def\makeblankbox#1#2{\hbox{\lower\dp0\vbox{\hidehrule{#1}{#2}%
   \kern -#1% overlap rules
   \hbox to \wd0{\hidevrule{#1}{#2}%
      \raise\ht0\vbox to #1{}% vrule height
      \lower\dp0\vtop to #1{}% vrule depth
      \hfil\hidevrule{#2}{#1}}%
   \kern-#1\hidehrule{#2}{#1}}}%
}%
\def\hidehrule#1#2{\kern-#1\hrule height#1 depth#2 \kern-#2}%
\def\hidevrule#1#2{\kern-#1{\dimen0=#1\advance\dimen0 by #2\vrule
    width\dimen0}\kern-#2}%
\def\openbox{\ht0=1.2mm \dp0=1.2mm \wd0=2.4mm  \raise 2.75pt
\makeblankbox {.25pt} {.25pt}  }

\def\bun#1/#2{\leavevmode
   \kern.1em \raise .5ex \hbox{\the\scriptfont0 #1}%
   \kern-.1em $/$%
   \kern-.15em \lower .25ex \hbox{\the\scriptfont0 #2}%
}

\def\opensquare{\ht0=3.4mm \dp0=3.4mm \wd0=6.8mm  \raise 2.7pt
\makeblankbox {.25pt} {.25pt}  }

%%%%%%%%%%%%%%%%%%%%%%%

\def\sector#1#2{\ {\scriptstyle #1}\hskip 1mm
\mathop{\opensquare}\limits_{\lower 1mm\hbox{$\scriptstyle#2$}}\hskip 1mm}

\def\tsector#1#2{\ {\scriptstyle #1}\hskip 1mm
\mathop{\opensquare}\limits_{\lower 1mm\hbox{$\scriptstyle#2$}}^\sim\hskip 1mm}
%%%
%%%

%% ANOTHER SET OF MACROS

\def\inbar{\,\vrule height1.5ex width.4pt depth0pt}

\font\cmss=cmss10 \font\cmsss=cmss10 at 7pt
\def\IR{\relax{\rm I\kern-.18em R}}

%% new macros

\def\frac#1#2{{#1\over#2}}

\def\half{\frac12}

\def\inbar{\,\vrule height1.5ex width.4pt depth0pt}
\def\IC{\relax\hbox{$\inbar\kern-.3em{\rm C}$}}
\def\IR{\relax{\rm I\kern-.18em R}}
\def\IP{\relax{\rm I\kern-.18em P}}

%
%%%%%%%%%%%%%%%%%%%%%%%%%%%%%%%%%%%%
%
\catcode`\@=11
\def\slash#1{\mathord{\mathpalette\c@ncel{#1}}}
\overfullrule=0pt

\def\II{{\cal I}}

\def\underrel#1\over#2{\mathrel{\mathop{\kern\z@#1}\limits_{#2}}}

\catcode`\@=12

%%%%%%%%%%%%%%%%%%%%%%%%%%%%%%%%%%%%%%%%%%%%%%%%%%%%%%%%%%%%%%

%

\def\cosh{{\rm cosh}}

\def\exp{{\rm exp}}

%%%%%%%%%%%%%%%%%%%%%%%%%%%%%%%%%%%%%%%%%%%%%%%%%%%%%%%%%%%%%%
% new defs:

%%%%%%%%%%%%%%%%%%%%%%%%%%%%%%%%%%%%%%%%%%%%%%%%%%%%%%%%%%%%%%

\def\frac#1#2{{#1\over#2}}

\def\half{\frac12}

\def\inbar{\,\vrule height1.5ex width.4pt depth0pt}
\def\IC{\relax\hbox{$\inbar\kern-.3em{\rm C}$}}
\def\IR{\relax{\rm I\kern-.18em R}}
\def\IP{\relax{\rm I\kern-.18em P}}

%
%%%%%%%%%%%%%%%%%%%%%%%%%%%%%%%%%%%%
%

%
\catcode`\@=11
\def\slash#1{\mathord{\mathpalette\c@ncel{#1}}}
\overfullrule=0pt

\def\II{{\cal I}}

\def\underrel#1\over#2{\mathrel{\mathop{\kern\z@#1}\limits_{#2}}}

\catcode`\@=12

%%%%%%%%%%%%%%%%%%%%%%%%%%%%%%%%%%%%%%%%%%%%%%%%%%%%%%%%%%%%%%

%

\def \cosh{{\rm cosh}}

\def\exp{{\rm exp}}

%%%%%%%%%%%%%%%%%%%%%%%%%%%%%%%%%%%%%%%%%%%%%%%%%%%%%%%%%%%%%%
% new defs:

%%%%%%%%%%%%%%%%%%%%%%%%%%%%%%%%%%%%%%%%%%%%%%%%%%%%%%%%%%%%%%%%%%%%%%%%%%%%%%%%%%

%\SugawaraHMA
\lref\SugawaraHMA{
  Y.~Sugawara,
  ``“Analytic continuation” of $N=2$ minimal model,''
PTEP {\bf 2014}, no. 4, 043B02 (2014).
[arXiv:1311.4708 [hep-th]].
%%CITATION = arXiv:1311.4708%%
}

\lref\GIKthree{A.~Giveon, N.~Itzhaki and D.~Kutasov, to appear.}

%\BekensteinUR
\lref\BekensteinUR{
  J.~D.~Bekenstein,
  ``Black holes and entropy,''
Phys.\ Rev.\ D {\bf 7}, 2333 (1973).
}

\lref\HawkingSW{
  S.~W.~Hawking,
  ``Particle Creation by Black Holes,''
Commun.\ Math.\ Phys.\  {\bf 43}, 199 (1975), [Erratum-ibid.\  {\bf 46}, 206 (1976)].
}

%\MaldacenaHW
\lref\MaldacenaHW{
  J.~M.~Maldacena and H.~Ooguri,
  ``Strings in AdS(3) and SL(2,R) WZW model 1.: The Spectrum,''
J.\ Math.\ Phys.\  {\bf 42}, 2929 (2001).
[hep-th/0001053].
%%CITATION = hep-th/0001053%%
}

%\PolchinskiRR
\lref\PolchinskiRR{
  J.~Polchinski,
  ``String theory. Vol. 2: Superstring theory and beyond,'' section 10.4.
}

%\BarbonNW
\lref\BarbonNW{
  J.~L.~F.~Barbon and E.~Rabinovici,
  ``Remarks on black hole instabilities and closed string tachyons,''
Found.\ Phys.\  {\bf 33}, 145 (2003).
[hep-th/0211212].
%%CITATION = CERN-TH-2002-313%%
}

%\ArgurioTB
\lref\ArgurioTB{
  R.~Argurio, A.~Giveon and A.~Shomer,
  ``Superstrings on AdS(3) and symmetric products,''
JHEP {\bf 0012}, 003 (2000).
[hep-th/0009242].
%%CITATION = hep-th/0009242%%
}

%\GiveonWN
\lref\GiveonWN{
  A.~Giveon, A.~Konechny, A.~Pakman and A.~Sever,
  ``Type 0 strings in a 2-d black hole,''
JHEP {\bf 0310}, 025 (2003).
[hep-th/0309056].
%%CITATION = hep-th/0309056%%
}

%\KutasovRR
\lref\KutasovRR{
  D.~Kutasov,
  ``Accelerating branes and the string/black hole transition,''
[hep-th/0509170].
%%CITATION = hep-th/0509170%%
}

\lref\fzz{
V.A. Fateev, A.B.
Zamolodchikov and Al.B. Zamolodchikov, unpublished.
}

%\KazakovPM
\lref\KazakovPM{
  V.~Kazakov, I.~K.~Kostov and D.~Kutasov,
  ``A Matrix model for the two-dimensional black hole,''
Nucl.\ Phys.\ B {\bf 622}, 141 (2002).
[hep-th/0101011].
%%CITATION = hep-th/0101011%%
}

%\GiveonPX
\lref\GiveonPX{
  A.~Giveon and D.~Kutasov,
  ``Little string theory in a double scaling limit,''
JHEP {\bf 9910}, 034 (1999).
[hep-th/9909110].
%%CITATION = hep-th/9909110%%
}

%%%%% Works on the importance of stringy drama in SL(2)/U(1) %%%%%%%%%%%

%\GiveonKP
\lref\GiveonKP{
  A.~Giveon and N.~Itzhaki,
  ``String Theory Versus Black Hole Complementarity,''
JHEP {\bf 1212}, 094 (2012).
[arXiv:1208.3930 [hep-th]].
%%CITATION = arXiv:1208.3930%%
}

%\GiveonICA
\lref\GiveonICA{
  A.~Giveon and N.~Itzhaki,
  ``String theory at the tip of the cigar,''
JHEP {\bf 1309}, 079 (2013).
[arXiv:1305.4799 [hep-th]].
%%CITATION = arXiv:1305.4799%%
}

%\MertensPZA
\lref\MertensPZA{
  T.~G.~Mertens, H.~Verschelde and V.~I.~Zakharov,
  ``Near-Hagedorn Thermodynamics and Random Walks: a General Formalism in Curved Backgrounds,''
JHEP {\bf 1402}, 127 (2014).
[arXiv:1305.7443 [hep-th]].
%%CITATION = arXiv:1305.7443%%
}

%\MertensZYA
\lref\MertensZYA{
  T.~G.~Mertens, H.~Verschelde and V.~I.~Zakharov,
  ``Random Walks in Rindler Spacetime and String Theory at the Tip of the Cigar,''
JHEP {\bf 1403}, 086 (2014).
[arXiv:1307.3491 [hep-th]].
%%CITATION = arXiv:1307.3491%%
}

%\GiveonHSA
\lref\GiveonHSA{
  A.~Giveon, N.~Itzhaki and J.~Troost,
  ``The Black Hole Interior and a Curious Sum Rule,''
JHEP {\bf 1403}, 063 (2014).
[arXiv:1311.5189 [hep-th]].
%%CITATION = arXiv:1311.5189%%
}

%\GiveonHFA
\lref\GiveonHFA{
  A.~Giveon, N.~Itzhaki and J.~Troost,
  ``Lessons on Black Holes from the Elliptic Genus,''
JHEP {\bf 1404}, 160 (2014).
[arXiv:1401.3104 [hep-th]].
%%CITATION = arXiv:1401.3104%%
}

%\MertensNCA
\lref\MertensNCA{
  T.~G.~Mertens, H.~Verschelde and V.~I.~Zakharov,
  ``The thermal scalar and random walks in $AdS_3$ and $BTZ$,''
JHEP {\bf 1406}, 156 (2014).
[arXiv:1402.2808 [hep-th]].
%%CITATION = arXiv:1402.2808%%
}

%\MertensCIA
\lref\MertensCIA{
  T.~G.~Mertens, H.~Verschelde and V.~I.~Zakharov,
  ``Near-Hagedorn Thermodynamics and Random Walks - Extensions and Examples,''
JHEP {\bf 1411}, 107 (2014).
[arXiv:1408.6999 [hep-th]].
%%CITATION = arXiv:1408.6999%%
}

%\WakimotoGF
\lref\WakimotoGF{
  M.~Wakimoto,
  ``Fock representations of the affine lie algebra A1(1),''
Commun.\ Math.\ Phys.\  {\bf 104}, 605 (1986).
}

%\BernardIY
\lref\BernardIY{
  D.~Bernard and G.~Felder,
  ``Fock Representations and BRST Cohomology in SL(2) Current Algebra,''
Commun.\ Math.\ Phys.\  {\bf 127}, 145 (1990).
%%CITATION = SACLAY-SPH-T-89-113%%
}

%\GiveonNS
\lref\GiveonNS{
  A.~Giveon, D.~Kutasov and N.~Seiberg,
  ``Comments on string theory on AdS(3),''
Adv.\ Theor.\ Math.\ Phys.\  {\bf 2}, 733 (1998).
[hep-th/9806194].
%%CITATION = hep-th/9806194%%
}

%\MertensDIA
\lref\MertensDIA{
  T.~G.~Mertens, H.~Verschelde and V.~I.~Zakharov,
  ``On the Relevance of the Thermal Scalar,''
JHEP {\bf 1411}, 157 (2014).
[arXiv:1408.7012 [hep-th]].
%%CITATION = arXiv:1408.7012%%
}

%\MertensSAA
\lref\MertensSAA{
  T.~G.~Mertens, H.~Verschelde and V.~I.~Zakharov,
  %``Perturbative String Thermodynamics near Black Hole Horizons,''
JHEP {\bf 1506}, 167 (2015).
[arXiv:1410.8009 [hep-th]].
%%CITATION = arXiv:1410.8009%%
}

%\GiveonCMA
\lref\GiveonCMA{
  A.~Giveon, N.~Itzhaki and D.~Kutasov,
  ``Stringy Horizons,''
JHEP {\bf 1506}, 064 (2015).
[arXiv:1502.03633 [hep-th]].
%%CITATION = arXiv:1502.03633%%
}

%\GiribetKCA
\lref\GiribetKCA{
  G.~Giribet and A.~Ranjbar,
  ``Screening Stringy Horizons,''
Eur.\ Phys.\ J.\ C {\bf 75}, no. 10, 490 (2015).
[arXiv:1504.05044 [hep-th]].
%%CITATION = arXiv:1504.05044%%
}

%\MertensHIA
\lref\MertensHIA{
  T.~G.~Mertens, H.~Verschelde and V.~I.~Zakharov,
  ``The long string at the stretched horizon and the entropy of large non-extremal black holes,''
JHEP {\bf 1602}, 041 (2016).
[arXiv:1505.04025 [hep-th]].
%%CITATION = arXiv:1505.04025%%
}

%\Ben-IsraelMDA
\lref\BenIsraelMDA{
  R.~Ben-Israel, A.~Giveon, N.~Itzhaki and L.~Liram,
  ``Stringy Horizons and UV/IR Mixing,''
JHEP {\bf 1511}, 164 (2015).
[arXiv:1506.07323 [hep-th]].
%%CITATION = arXiv:1506.07323%%
}

%\MertensOLA
\lref\MertensOLA{
  T.~G.~Mertens,
  ``Hagedorn String Thermodynamics in Curved Spacetimes and near Black Hole Horizons,''
[arXiv:1506.07798 [hep-th]].
%%CITATION = arXiv:1506.07798%%
}

%\LinLFA
\lref\LinLFA{
  J.~Lin,
  ``Bulk Locality from Entanglement in Gauge/Gravity Duality,''
[arXiv:1510.02367 [hep-th]].
%%CITATION = arXiv:1510.02367%%
}

%\Ben-IsraelETG
\lref\BenIsraelETG{
  R.~Ben-Israel, A.~Giveon, N.~Itzhaki and L.~Liram,
  ``On the Stringy Hartle-Hawking State,''
JHEP {\bf 1603}, 019 (2016).
[arXiv:1512.01554 [hep-th]].
%%CITATION = arXiv:1512.01554%%
}

%\CottrellNSU
\lref\CottrellNSU{
  W.~Cottrell and A.~Hashimoto,
  ``Resolved gravity duals of ${\cal N}=4$ quiver field theories in 2+1 dimensions,''
[arXiv:1602.04765 [hep-th]].
%%CITATION = MAD-TH-02%%
}

%%%%%%%%%%%%%%% End of stringy drama in SL(2)/U(1) %%%%%%%%%%%%%%%%%%%%%%%%%%%%%%

%%%%%%%%%%% 2d BH %%%%%%%%%%%%%%%%%%%

%\ElitzurCB
\lref\ElitzurCB{
  S.~Elitzur, A.~Forge and E.~Rabinovici,
  ``Some global aspects of string compactifications,''
Nucl.\ Phys.\ B {\bf 359}, 581 (1991).
%%CITATION = RI-143-90%%
}

%\MandalTZ
\lref\MandalTZ{
  G.~Mandal, A.~M.~Sengupta and S.~R.~Wadia,
  ``Classical solutions of two-dimensional string theory,''
Mod.\ Phys.\ Lett.\ A {\bf 6}, 1685 (1991).
%%CITATION = IASSNS-HEP-91-10%%
}

%\WittenYR
\lref\WittenYR{
  E.~Witten,
  ``On string theory and black holes,''
Phys.\ Rev.\ D {\bf 44}, 314 (1991).
%%CITATION = IASSNS-HEP-91-12%%
}

%\DijkgraafBA
\lref\DijkgraafBA{
  R.~Dijkgraaf, H.~L.~Verlinde and E.~P.~Verlinde,
  ``String propagation in a black hole geometry,''
Nucl.\ Phys.\ B {\bf 371}, 269 (1992).
%%CITATION = PUPT-1252%%
}

%%%%%%%%%%% end of 2d BH %%%%%%%%%%%%

%\KazamaQP
\lref\KazamaQP{
  Y.~Kazama and H.~Suzuki,
 ``New N=2 Superconformal Field Theories and Superstring Compactification,''
Nucl.\ Phys.\ B {\bf 321}, 232 (1989).
%%CITATION = UT-KOMABA-88-8%%
}

%\BershadskyIN
\lref\BershadskyIN{
  M.~Bershadsky and D.~Kutasov,
  ``Comment on gauged WZW theory,''
Phys.\ Lett.\ B {\bf 266}, 345 (1991).
%%CITATION = PUPT-1261%%
}

%\HorowitzNW
\lref\HorowitzNW{
  G.~T.~Horowitz and J.~Polchinski,
  ``A Correspondence principle for black holes and strings,''
Phys.\ Rev.\ D {\bf 55}, 6189 (1997).
[hep-th/9612146].
%%CITATION = hep-th/9612146%%
}

%\HorowitzJC
\lref\HorowitzJC{
  G.~T.~Horowitz and J.~Polchinski,
  ``Selfgravitating fundamental strings,''
Phys.\ Rev.\ D {\bf 57}, 2557 (1998).
[hep-th/9707170].
%%CITATION = hep-th/9707170%%
}

%\GiveonPR
\lref\GiveonPR{
  A.~Giveon and D.~Kutasov,
  ``Fundamental strings and black holes,''
JHEP {\bf 0701}, 071 (2007).
[hep-th/0611062].
%%CITATION = hep-th/0611062%%
}

%%%%%%%%%%%%%%%%%%%%%%%%%%%%%%%%%%%%%%%%%%%%%%%%%%%%%%%%%%%%%%%%%%%%%%%%%%%%%
%\rightline{....}
\Title{
%\rightline{hep-th/yymmnnn}
} {\vbox{
\bigskip\centerline{Stringy Horizons II}}}
\medskip
\centerline{\it Amit Giveon${}^{1}$, Nissan Itzhaki${}^{2}$ and David Kutasov${}^{3}$}
\bigskip
\smallskip
\centerline{${}^{1}$Racah Institute of Physics, The Hebrew
University} \centerline{Jerusalem 91904, Israel}
\smallskip
\centerline{${}^{2}$ Physics Department, Tel-Aviv University, Israel} \centerline{Ramat-Aviv, 69978, Israel}
\smallskip
\centerline{${}^3$EFI and Department of Physics, University of
Chicago} \centerline{5640 S. Ellis Av., Chicago, IL 60637, USA }

\bigskip\bigskip\bigskip
\noindent

We show that the spectrum of normalizable states on a Euclidean $SL(2,R)/U(1)$ black hole exhibits a duality between oscillator states  and wound strings. This duality generalizes the FZZ correspondence, which can be thought of as an identification between a normalizable mode of dilaton gravity and a mode of the tachyon with winding number one around the Euclidean time circle. It implies that normalizable states on a large Euclidean black hole have support at widely separated scales. In particular, localized states that are extended over the cap of the cigar (the Euclidian analog of the black hole atmosphere) have a component that is localized near the tip of the cigar (the analog of the stretched horizon). As a consequence of this duality, the states exhibit a transition as a function of radial excitation level. From the perspective of a low energy probe, low lying states are naturally thought of as oscillator states in the black hole atmosphere, while at large excitation level they are naturally described as wound strings. As the excitation level increases, the size of the states first decreases and then increases. This behavior is expected to be a general feature of black hole horizons in string theory.

\vglue .3cm
%\vskip 2cm
\bigskip

\Date{3/16}

\bigskip

\newsec{Introduction}

One of the important open problems in quantum gravity is the origin of the Bekenstein-Hawking entropy of black holes~\refs{\BekensteinUR,\HawkingSW}.
In particular, it is still not clear where the states responsible for the black hole entropy are located. A priori, one might expect them to reside inside the black hole, on the horizon, or in the thermal atmosphere. There are, however, difficulties with all these options.
In this note we present arguments that indicate that  string theory might shed light on this question.

String theory modifies classical gravity in two ways. There are string effects, whose typical scale is the string scale $l_s$, and quantum effects, whose scale is the Planck scale $l_p$.  In weakly coupled string theory the hierarchy of scales is $l_s\gg l_p$, and it is natural to ask whether string theory modifies the picture obtained in classical gravity already at the scale $l_s$, well above the Planck (length) scale. This question can be studied in classical string theory, by including $\alpha' (=l_s^2)$ effects.

One may hope to get information about the physics associated with the horizon of a black hole by Wick rotating to Euclidean spacetime. The advantage of this continuation is that it gives rise to a smooth geometry, with the radial and Euclidean time direction forming a semi-infinite cigar geometry; the tip of the cigar is the Euclidean continuation of the horizon of the black hole.

When studying string propagation in this background, the winding of the string around the Euclidean time direction is not conserved, since the string can slip off the tip of the cigar. It turns out that there is another source of winding non-conservation -- a condensate of the winding tachyon field, the lowest mode of a string winding around the Euclidean time circle at infinity, which is necessarily present in Euclidean black hole spacetimes \KutasovRR\ (see also \BarbonNW).  For large black holes, one can think of the tachyon condensate as a non-perturbative $\alpha'$ effect.

The two mechanisms for string winding violation mentioned in the previous paragraph are superficially different, but it is believed that both are present and are not independent \KutasovRR. In particular, the size of the winding tachyon condensate is determined by the geometry of the Euclidean black hole.

While the above picture is expected to be general, it's been studied in detail primarily for a particular case --  the two dimensional black hole corresponding to the coset conformal field theory (CFT)
$SL(2,\IR)/U(1)$~\refs{\ElitzurCB\MandalTZ\WittenYR-\DijkgraafBA}.
The latter is exactly solvable due to its relation to the CFT on the $SL(2,\IR)$ group manifold, so one can study its physics in detail. In particular, one can ask the question what are the implications of the $\alpha'$ effects on the questions mentioned above
\refs{\KutasovRR,\GiveonKP\GiveonICA\MertensPZA\MertensZYA\GiveonHSA\GiveonHFA
\MertensNCA\MertensCIA\MertensDIA\MertensSAA\GiveonCMA\GiribetKCA\MertensHIA
\BenIsraelMDA\MertensOLA\LinLFA\BenIsraelETG-\CottrellNSU}.
In this note we will continue our study of this question.

The existence of the winding tachyon condensate in the two dimensional black hole background is known as the FZZ correspondence \fzz; see \KazakovPM\ for a review.  There is an analog of this correspondence in the theory with $\CN=2$ worldsheet supersymmetry \GiveonPX, which plays a role in the superstring. Here we will mostly discuss the bosonic case, and comment briefly on the supersymmetric generalization towards the end.

In the original work on the FZZ correspondence \refs{\fzz,\KazakovPM}, it was thought of as a duality between the CFT's describing large and small black holes. It was later realized that this correspondence plays a role in the physics of large black holes as well
\refs{\KutasovRR,\GiveonKP\GiveonICA\MertensPZA\MertensZYA\GiveonHSA\GiveonHFA
\MertensNCA\MertensCIA\MertensDIA\MertensSAA\GiveonCMA\GiribetKCA\MertensHIA
\BenIsraelMDA\MertensOLA\LinLFA\BenIsraelETG-\CottrellNSU}.
In particular, \KutasovRR\ argued that the tachyon condensate gives rise to a smearing of the horizon of a black hole. In \GiveonCMA\ it was shown that scattering particles off the tip of the cigar gives rise to an interesting effect. While low energy particles scatter in the cigar geometry in the way dictated by general relativity (GR), high energy ones do not see the tip of the cigar and instead are sensitive to the winding tachyon condensate. This gives a scattering phase shift which grows with energy, in sharp contrast to GR, where the phase shift goes to a constant at high energy due to the fact that space ends at the tip of the cigar. Thus, one can say that the FZZ correspondence is a high/low energy duality.

One can also think of the FZZ correspondence as an identification of two seemingly different normalizable modes on the cigar. One governs the value of the dilaton at the tip of the cigar, or the metric deformation that closes up the infinite cylinder to a cigar. The other is the winding tachyon. The two modes have in general different behaviors at infinity -- the former is much more extended (in the radial direction) than the latter for large black holes. Nevertheless, the FZZ correspondence ties the two.

We will see here that the identification between the dilaton and the wound tachyon is a special case of a more general phenomenon, which we will refer to as the  {\it generalized FZZ correspondence}.  This correspondence relates a large class of seemingly distinct normalizable states on the cigar, which behave in a different way in the asymptotic region. It can also be thought of as a high/low energy correspondence.
%, but in a different sense than that of \GiveonCMA.

The plan of this note is the following. In section 2 we briefly review the geometry of the Euclidean $SL(2,\IR)/U(1)$ black hole, and describe a class of normalizable states in this background. In section 3 we review the description of (part of) the spectrum of normalizable states on the cigar in terms of strings winding around the cigar, that experience an attractive potential towards the tip. We point out that a semiclassical analysis of the bound state problem for this potential gives the correct energies for these states, but that one expects the detailed properties of these states to receive large corrections at low excitation levels.

These corrections are governed by the generalized FZZ (GFZZ) correspondence, which is described for pure winding states in section 4. We start this section by describing the spectrum of normalizable states in the CFT on $SL(2,\IR)$, and in particular review some properties of  the principal discrete series representations, the representations obtained from them via spectral flow, and the isomorphism between the two. We then use this isomorphism to derive an identification between seemingly different states on the cigar, relating a large class of states with winding numbers zero and one.

In section 5 we discuss the asymptotic form of the vertex operators describing a particular class of GFZZ dual states. We start by describing their ancestors in the underlying $SL(2,\IR)$ CFT using the Wakimoto representation, which is useful for studying the theory near the boundary of $AdS_3$. We then present an approach to finding the vertex operators in the coset from those in $AdS_3$. We find that the GFZZ correspondence relates in this case oscillator states with winding number zero to states with winding number one and oscillator number zero.

In section 6 we discuss the physical consequences of the GFZZ correspondence. Section 7 is devoted to some generalizations of the correspondence. Finally, in section 8 we summarize our results and comment on their possible implications for Lorentzian black holes.

\newsec{Some properties of the two dimensional Euclidean black hole}

The Euclidean $SL(2,\IR)/U(1)$ coset CFT describes string propagation on a semi-infinite cigar~\refs{\ElitzurCB\MandalTZ\WittenYR-\DijkgraafBA}, with metric and dilaton\foot{We present the background to leading order in $1/k$, and  have chosen the convention $\alpha'=2$.}
\eqn\sltwo{\eqalign{&ds^2=2k(dr^2+\tanh^2rd\theta^2)~;\cr
&\Phi-\Phi_0=-\ln\cosh r~.}}
$\theta\sim \theta+2\pi$ is an angular coordinate, obtained by Wick rotating the time coordinate. The radial coordinate $0\le r<\infty$ is the direction along the cigar; $r=0$ is the tip, while for large $r$ (compared to $1$) the background \sltwo\ approaches a cylinder of radius $\sqrt{2k}$, with linear dilaton along it. The string coupling $e^\Phi$ depends on $r$; it goes to zero far from the tip and attains its maximal value, $e^{\Phi_0}$, at the tip. This value controls the mass of the black hole. The region of size of order $\sqrt{k}$ around the tip ($r$ of order $1$ in \sltwo) is the cap of the cigar (see figure 1); in this region the curvature is of order $1/k$.

\ifig\loc{The tip (denoted in red) and cap of the cigar.}
{\epsfxsize2.5in\epsfbox{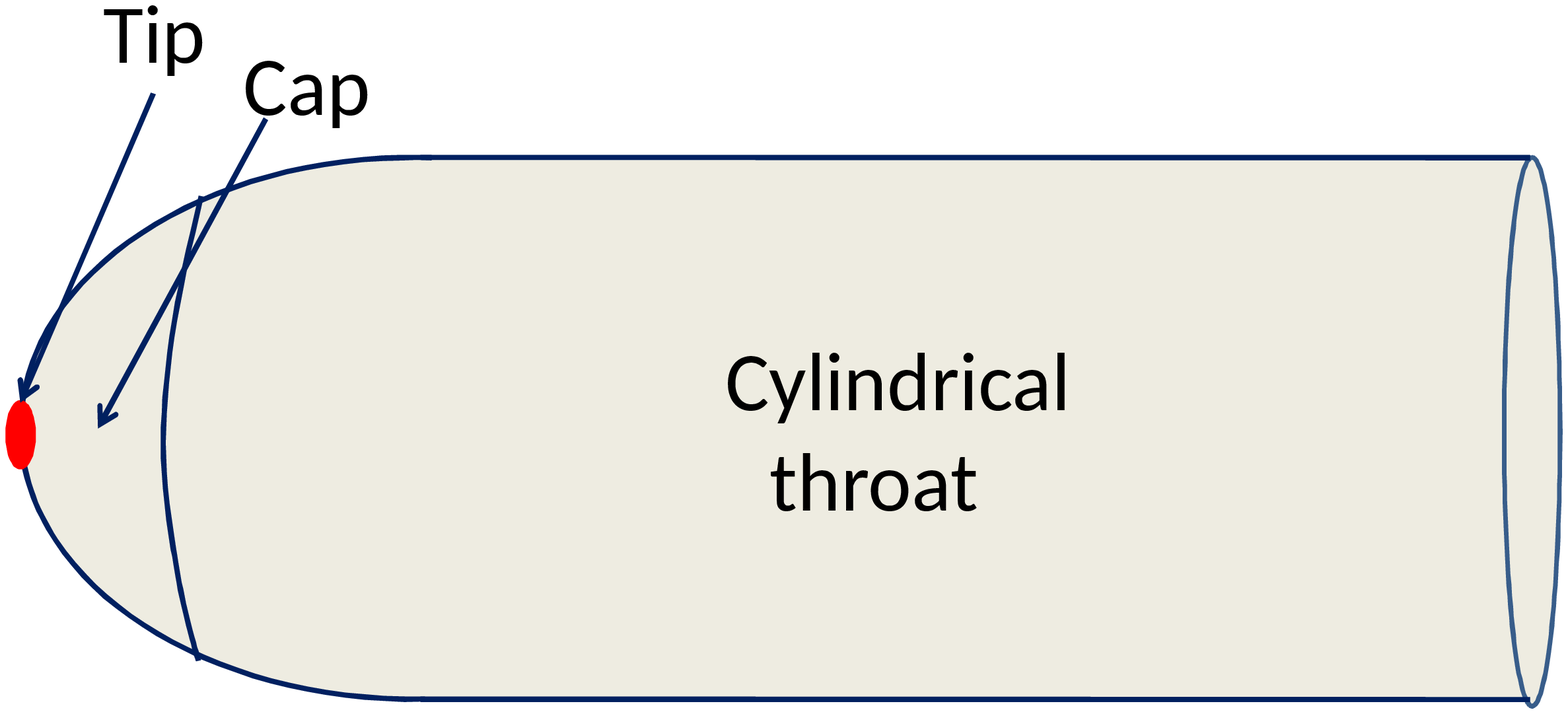}}

$k$ is a free parameter, which governs the overall size of the cigar. In the algebraic coset description, it corresponds to the level of the underlying $SL(2,\IR)$ current algebra. Geometrically, it sets the overall scale of the cigar. In particular, at large $k$ \sltwo\ describes a weakly curved geometry. That is the analog in this context of a large (Euclidean) Schwarzschild black hole in higher dimensions.

The model comes in two versions, depending on whether one is studying it in the bosonic string or the superstring. In the former case, one is interested in the bosonic coset model, whose central charge is given by
\eqn\bosc{c=2+{6\over k-2}~.}
The background fields \sltwo\ receive perturbative $\alpha'$ corrections
\DijkgraafBA, which can be thought of as $1/k$ corrections.

In the superstring, one is interested in the $\CN=1$ superconformal coset,\foot{Which happens to have $\CN=2$ superconformal symmetry; this is an example of the Kazama-Suzuki \KazamaQP\ construction.} which is obtained by attaching to a bosonic $SL(2,\IR)$ WZW model three free fermions that transform in the adjoint representation of $SL(2,\IR)$, and gauging the diagonal $U(1)$ in the full $SL(2,\IR)$ of bosons $+$ fermions. The total level of $SL(2,\IR)$, $k$, can be written in this case as a sum of bosonic and fermionic contributions, $k=(k+2)+(-2)$, and the corresponding central charge is
\eqn\supc{c=3+{6\over k}~.}
In this case, the background \sltwo\ does not receive perturbative corrections in $1/k$.

Although, as usual in string theory, to talk about a well defined theory with a stable vacuum one needs to consider the superstring, for our purposes the bosonic theory is good enough, since the physics we are interested in is unrelated to the usual closed string tachyon. Hence, we will phrase the discussion below in this language; we will comment briefly on the worldsheet supersymmetric case in section 7.\foot{The details will appear
in a separate work \GIKthree.}

We will be interested here in normalizable states on the cigar. A large class of such states is described by the Virasoro primary vertex operators $V_{j;m,\bar m}$, whose scaling dimensions are given by
\eqn\delbar{\eqalign{
\Delta_{j;m,\bar m}&=-{j(j+1)\over k-2}+{m^2\over k}~,\cr
\bar\Delta_{j;m,\bar m}&=-{j(j+1)\over k-2}+{\bar m^2\over k}~.
}}
Here, $m$ and $\bar m$ label momentum and winding around the cigar; they take the values
\eqn\momwin{\eqalign{m=&\half(wk-p)~,\cr
                                  \bar m=&\half(wk+p)~,
                                  }}
where $p,w\in\IZ$ are the momentum and winding around the circle labeled by $\theta$, respectively. Note that while the momentum on the circle $p$ is conserved, the winding $w$ is not, as winding can slip off the tip of the cigar.

The quantum number $j$ governs the  radial dependence of the wavefunctions of the states \delbar. It takes value in the range
\eqn\normstat{j=|m|-n=|\bar m|-\bar n~, \qquad n,\bar n=1,2,3,\cdots~.}
Unitarity of the CFT leads to a bound on $j$,
\eqn\unitbd{-{1\over2}<j<{k-3\over2}~,}
which in turn implies a bound on the integers $(n,\bar n)$ in \normstat.

The states \delbar\ -- \normstat\ are known as principal discrete series states, since they descend in the GKO coset construction from  analogous states in CFT on the $SL(2,\IR)$ group manifold. The vertex operators that create them, $V_{j;m,\bar m}$, behave far from the tip of the cigar as
\eqn\behavvv{V_{j;m,\bar m}\simeq e^{ip_LX_L+ip_RX_R-Q(j+1)\phi}~,}
where $(\phi,X)$ are canonically normalized fields, in terms of which the metric \sltwo\ behaves at infinity like $ds^2=d\phi^2+dX^2$, and the dilaton goes like $\Phi=-{Q\over2}\phi$. The background charge $Q$ is related to $k$ via
\eqn\qqq{Q=\sqrt{2\over k-2}~;}
for large black holes (large $k$) it goes to zero like $Q\sim\sqrt{2\over k}$. The left and right-moving momenta on a circle of radius $R$, $p_L={p\over R}-{wR\over2}$, $p_R={p\over R}+{wR\over2}$, are related to $(m,\bar m)$  \momwin\ via the relation
\eqn\plpr{(p_L,p_R)=\sqrt{2\over k}(-m,\bar m)~.}
As mentioned above, the radius of the circle is $R=\sqrt{2k}$.

\newsec{Semiclassical description of states with $w=1$, $p=0$}

To introduce the issue we will focus on, consider the states \delbar\ with $w=1$, $p=0$. Looking back at \momwin, we see that in this case $m=\bar m=k/2$, and \normstat, \unitbd\  imply that $j$ takes the values\foot{Here we assume that ${k+1\over2}\not\in\IZ$. If it is, the upper bound is smaller by one.}
\eqn\valjjj{j={k\over2}-n~;\qquad n=2,3,\cdots, \left[k+1\over2\right]~.}
The dimensions \delbar\ take in this case the form
\eqn\dimpurewin{\Delta_n=\bar\Delta_n={(n-1)(k-n)\over k-2}~.}
As $n$ varies over the range \valjjj, the dimension \dimpurewin\ varies between $1$ for $n=2$ and a value of order $k/4$ (for large $k$) for the largest value of $n$.

One can attempt to understand this spectrum qualitatively by studying the dynamics of a wound string in the cigar geometry \sltwo. The energy of a string winding the $\theta$ circle depends on its radial position. It goes to zero as $r\to 0$ (where the string can unwind), and monotonically increases with $r$, approaching a constant at large $r$. Thus, the radial equation for the zero mode of the string looks like a Schroedinger equation in a potential which has the above qualitative structure.

This equation was studied in \refs{\DijkgraafBA,\MertensZYA,\GiveonHFA} (for general $p$). It takes the form
\eqn\aaaaaa{
(L_0 +\bar{L}_0) |\Psi \rangle= (\Delta+\bar{\Delta}) |\Psi \rangle~,}
with
\eqn\op{
L_0=-{\triangle(r)\over k-2}+{m^2\over k}~,~~~~
\bar{L}_0=-{\triangle(r)\over k-2}+{\bar m^2\over k}~,
}
where $\triangle(r)$ is the Laplacian  on $SL(2,\IR)$.\foot{$\triangle(r)$ is given \eg\ in eq. (29) of \GiveonHFA, with $k\to k-2$ and $\rho=2r/Q$.} The eigenvalues $\Delta,\bar{\Delta}$ that one finds by solving  \aaaaaa, \op\ are precisely those given in \delbar, \momwin. Moreover, the eigenfunctions are known exactly \MertensZYA;\foot{See, in particular, appendices E and F.} their asymptotic behavior agrees with  \behavvv.

The quantum number $n$ \valjjj\ can be thought of as the radial excitation level. The lowest state has $n=2$, and a wavefunction that is highly localized in the radial direction, corresponding to \behavvv\ with $j={k\over2}-2$. It decays at large $\phi$ as $\exp(-\phi/Q)$, and corresponds to the Sine-Liouville vertex operator \KazakovPM\ that is localized at the tip of the cigar. As $n$ increases, $j$ \valjjj\ decreases; the corresponding vertex operator \behavvv\ becomes more spread out in the radial direction. As $n$ approaches the upper bound of the range \valjjj, the wavefunction spreads over a larger and larger part of the cigar. In that region one has
\eqn\largenn{n={k+1\over 2}-\alpha~,}
with $\alpha$ an order one (in the sense of the $1/k$ expansion) positive real number, and the vertex operator \behavvv\ decays at large $\phi$ as
\eqn\decayv{e^{-Q(\half+\alpha)\phi}~.}
Stripping off the factor of the string coupling $g_s\sim\exp(-Q\phi/2)$ that relates the vertex operator to the wavefunction, we find that the wavefunction of the state behaves at large $\phi$ as $\exp(-\alpha Q\phi)$, \ie\ it extends over a finite fraction of the cap of the cigar.

The description of the bound states \aaaaaa, \op\ is obtained by considering a straight string wrapping $\theta$ (by working in the gauge  $\theta(\sigma)=\sigma$, where $\sigma$ is the spacelike worldsheet coordinate)  and studying its radial dynamics. As mentioned above, this description gives the correct values of the scaling dimensions \delbar. It is natural to ask whether it also gives a correct description of the detailed structure of these states, particularly for large $k$, when the cigar \sltwo\ is large and weakly curved, and the dilaton is slowly varying.

Superficially, one would say that the winding description should only be valid when the wavefunction of the bound state is supported primarily in the region of large $\phi$, where the wound string is long and the semiclassical approximation is valid. This is the case for highly excited states, with $n$ towards the top of the range \valjjj. Such states can be characterized by the winding around the cigar, which is approximately conserved.

For low lying states, with $n$ close to the bottom of that range, one would expect large corrections to the picture \aaaaaa, \op. Indeed,  the semiclassical analysis gives in that case wavefunctions that are supported in the small $\phi$ region near the tip of the cigar, where the winding number is not conserved. In that region we expect to be able to describe the target space as an almost flat two dimensional space. Hence, the bound states should be related to standard perturbative string oscillator states, which seem very different from the straight strings with only radial oscillations described by \aaaaaa, \op.

In the rest of this note we will study this question in more detail. We will see that the low lying states are much more extended in the radial $(\phi)$ direction than implied by the analysis \aaaaaa, \op. Their large $\phi$ behavior is described in terms of oscillator states of a string in the weakly curved space \sltwo. This description is related to the one in terms of winding strings by a generalization of the FZZ correspondence. The semiclassical winding string description reviewed in this section can be neglected when studying long distance properties of these states, but plays an important role in analyzing the properties of these states sensitive to the region near the tip of the cigar.

Thus, the generalized FZZ correspondence is a high/low energy duality in two different senses. One is similar to the original FZZ duality: for low lying normalizable states, low energy probes see an oscillator state, while high energy probes see the winding string component of the wavefunction, as in \GiveonCMA. The other is that as one varies the excitation level $n$, for small $n$ low energy probes see an oscillator state, while for large $n$ they see a winding string.

\newsec{Generalized FZZ correspondence I: $p=0$ ($w=0$ vs. $w=1$)}

To make the picture described at the end of the previous section precise, it is convenient to use the description of the cigar CFT as the coset $SL(2,\IR)/U(1)$. Normalizable states on the cigar descend from normalizable states in the $SL(2,\IR)$ CFT. Therefore, we start by reviewing some properties of the latter \MaldacenaHW.

The  left-moving $SL(2,\IR)$ current algebra at level $k$ that governs the dynamics of this model is
\eqn\slalg{[J_n^3,J_m^3]=-{k\over 2}n\delta_{n+m,0}~,
\quad [J_n^3,J_m^\pm]=\pm J^\pm_{n+m}~,
\quad [J_n^+,J_m^-]=-2J^3_{n+m}+kn\delta_{n+m,0}~.}
There is a similar algebra for the right-movers; here and below we will often focus on the left-movers. The $SL(2,\IR)/U(1)$ coset CFT is obtained by modding out by the $U(1)$ current $J^3$ that generates a compact abelian subgroup of $SL(2,\IR)$. Therefore, it is useful to classify the states in the underlying $SL(2,\IR)$ CFT according to their $J^3$ eigenvalues.

The affine Lie algebra \slalg\ has two conjugate types of representations known as principal discrete series representations, $\hat D_j^{\pm,w=0}$, where $-\half<j\in\IR$. The lowest weight representation $\hat D_j^{+,w=0}$ is built on top of a lowest weight state $|j,m\rangle$. Here $m$ is the eigenvalue of $J^3_0$; it is related to $j$ by $m=j+1$. This state has dimension (eigenvalue of $L_0$) $\Delta_j=-j(j+1)/(k-2)$. It is annihilated by $J^-_0$; when acting on it with $J^+_0$, one finds states $|j,m\rangle$ with $m>j+1$ (and the same dimension). Acting with raising operators of the affine Lie algebra, $J^a_{-n}$ with $a=3,\pm$,  gives states with larger dimensions (current algebra descendants).  The conjugate representation $\hat D_j^{-,w=0}$ is obtained by flipping all the signs of the eigenvalues of $J^3_0$. In particular, it is built on top of a state $|j,m=-j-1\rangle$, which is annihilated by $J_0^+$. The $J^3_0$ and $L_0$ eigenvalues of the states in the representations $\hat D_j^{\pm,w=0}$ are depicted in figure 2.

\ifig\loc{The $\widehat{SL}(2,\IR)$ representations $\hat D_j^{+,w=0}$ (a), and $\hat D_j^{-,w=0}$ (b).}
{\epsfxsize4.5in\epsfbox{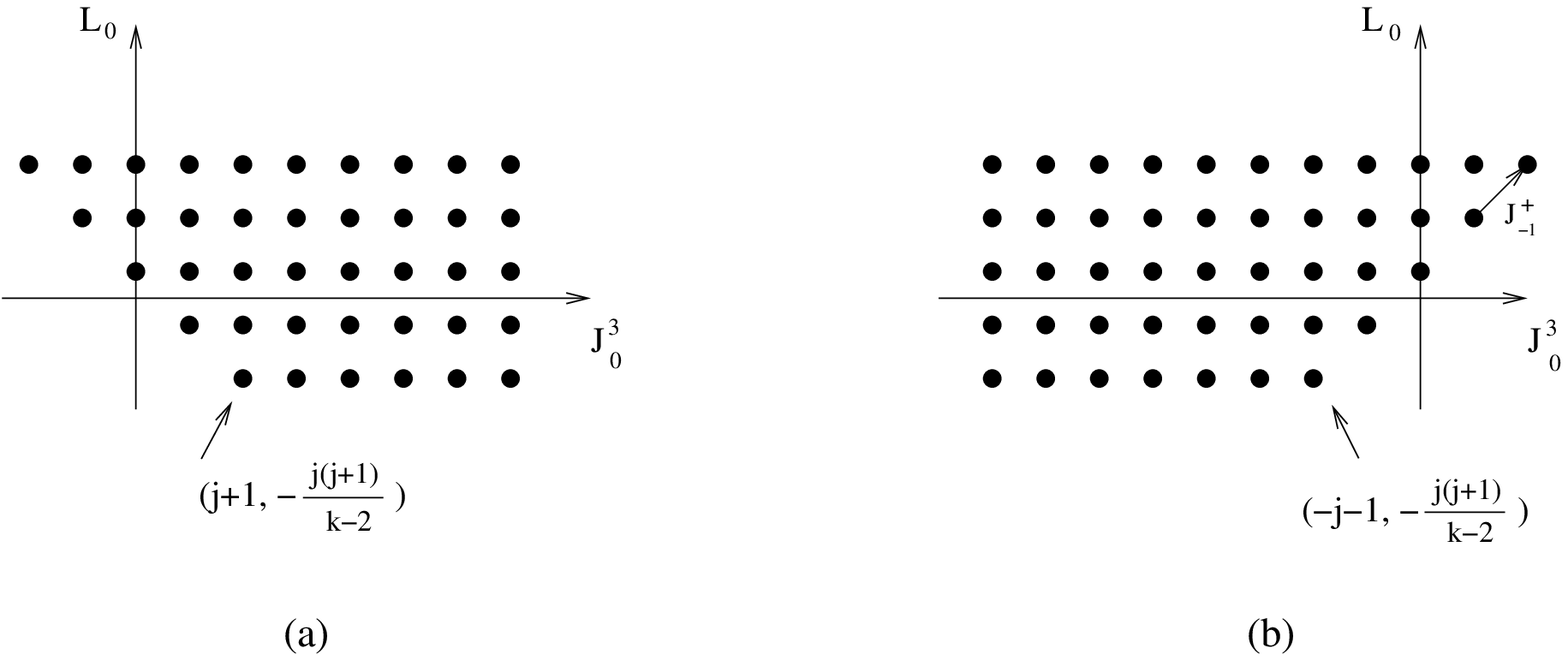}}

In addition to the principal discrete series states described in the previous paragraph, the theory has representations labeled by an integer $w$, which can be thought of as a winding number around the spatial direction on the boundary of $AdS_3$. This winding number is of course not conserved, since strings that wind around the boundary can shrink through the bulk of $AdS_3$.

Algebraically, the quantum number $w$ is  associated with an automorphism of the algebra \slalg\  known as spectral flow \MaldacenaHW,
\eqn\sflow{\tilde J_n^3=J_n^3+{k\over 2}w\delta_{n,0}~,\quad
\tilde J^\pm_n=J^\pm_{n\mp w}~,}
where $w\in\IZ$ is the spectral flow parameter. One can use this automorphism to define states $|j,m=j+1;w\rangle$, which belong to the representation $\hat D_j^{+,w}$. This representation is obtained by taking the representation $\hat D_j^{+,w=0}$ of the algebra $\tilde J^a_n$, and viewing it as a representation of the algebra $J^a_n$, related to $\tilde J^a_n$ via \sflow. The lowest weight state in the representation, $|j,m=j+1;w\rangle$, thus has $\tilde J_0^3=j+1$ and  \sflow\ $J^3_0=j+1-\half kw$; it is annihilated by $\tilde J_0^-=J_w^-$. The states $|j,m;w\rangle$ with $m=j+1+\ell$ are obtained  by acting on it $\ell$ times with $\tilde J_0^+=J_{-w}^+$; their
scaling dimension $\Delta_{j;m;w}$ and $J_0^3$ eigenvalue are
\eqn\dimjw{\Delta_{j;m;w}=-{j(j+1)\over k-2}+mw-{kw^2\over 4}~,
\qquad J_0^3=m-{k\over 2}w~.}
Other states in the representation $\hat D_j^{+,w}$ are obtained by acting on $|j,m;w\rangle$ with $\tilde J_{-n}^\pm=J_{-n\mp w}^\pm$ and $\tilde J^3_{-n}=J^3_{-n}$.

An important fact for our purposes is that the representations $\hat D_j^{-,w=0}$ and $\hat D^{+,w=1}_{\tilde j={k\over 2}-j-2}$ are isomorphic. In particular, the states $|j,m=-(j+1);w=0\rangle$ and $|\tilde j,m=\tilde j+1;w=1\rangle$ have the same values of $L_0$ and $J_0^3$,
\eqn\samelj{\eqalign{L_0=&-{j(j+1)\over k-2}~,\cr
J_0^3=&-j-1~,}}
 as can be checked directly by using  \dimjw. All other states in the two representation have the same quantum numbers as well, as can be seen in figure 3.

 \ifig\loc{The isomorphism between $\hat D_{\tilde j}^{+,w=1}$ and $\hat D_j^{-,w=0}$, with $\tilde j={k\over 2}-j-2$.}
{\epsfxsize2in\epsfbox{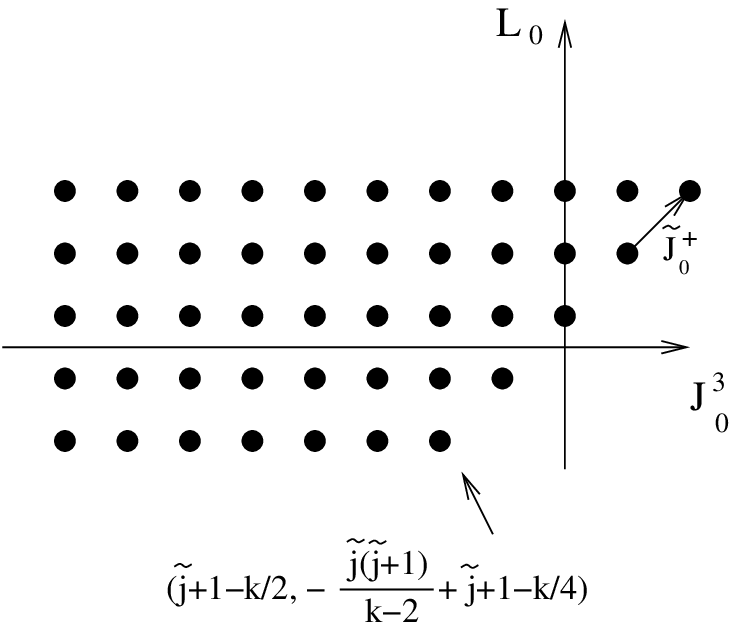}}

Since the representations $\hat D_j^{-,w=0}$ and $\hat D^{+,w=1}_{\tilde j={k\over 2}-j-2}$ are isomorphic, it is natural to ask whether they should be identified in CFT on $SL(2,\IR)$. We will assume that this is indeed the case, and will see that this assumption is consistent with some known facts about the coset $SL(2,\IR)/U(1)$. In principle, one should be able to show this directly in the $SL(2,\IR)$ CFT, but as far as we know this has not been done.

So far we discussed the situation in the CFT on $SL(2,\IR)$. We now turn to the coset CFT, and in particular to the question what the identification of $\hat D_j^{-,w=0}$ and $\hat D^{+,w=1}_{\tilde j={k\over 2}-j-2}$ in the former implies for the latter. It will be convenient to discuss this question in terms of the normalizable vertex operators that create the various states when acting on the vacuum. Thus, we start by reviewing their structure.

We begin with the vertex operators corresponding to the principal discrete series states $|j,m,\bar m;w=0\rangle$ in $SL(2,\IR)$ CFT, $\Phi_{j;m,\bar m}$, which have dimension $-j(j+1)/(k-2)$, and charge $m$ $(\bar m)$ under $J^3$ $(\bar J^3)$. We can write them as a product of contributions from $SL(2,\IR)/U(1)$ and the $U(1)$ CFT corresponding to $(J^3,\bar J^3)$ as follows. Using \slalg, we can write the currents  $(J^3,\bar J^3)$ as
\eqn\defxxx{J^3=-\sqrt{k\over 2}\partial x~;\qquad \bar J^3=-\sqrt{k\over 2}\bar\partial \bar x~,}
where $x$, $\bar x$ are canonically normalized left and right-moving scalars. The OPE
\eqn\opejphi{J^3(z)\Phi_{j;m,\bar m}(w)={m\over z-w}\Phi_{j;m,\bar m}(w)}
implies that we can write $\Phi_{j;m,\bar m}$ as
\eqn\decomphi{\Phi_{j;m,\bar m}=V_{j;m,\bar m}e^{{\sqrt{2\over k}(mx+\bar m \bar x)}}~,}
where $V_{j;m,\bar m}$ is a vertex operator that commutes with the charges $J^3_n$ \slalg; its dimension is given by \delbar. In general, the quantum numbers $(m,\bar m)$ do not take the values \momwin\ with integer $(p,w)$, so the operator $V_{j;m,\bar m}$ in \decomphi\ is not a good vertex operator in the coset. This is related to the fact that $SL(2,\IR)$ cannot be written as a direct product $SL(2,\IR)/U(1)\times U(1)$. However, by a judicious choice of the quantum numbers $(j;m,\bar m)$, one can construct vertex operators that do belong to the coset.

An example that will play a role in our discussion below is the state
\eqn\pzero{(J_{-1}^+)^l(\bar J_{-1}^+)^l|j=l-1;m=\bar m=-l;w=0\rangle~,}
with $l$ a positive integer. Before applying the raising operators $J_{-1}^+$, $\bar J_{-1}^+$, we have a primary state in the principal discrete series representation $\hat D_{l-1}^{-,w=0}$. The corresponding vertex operator is $\Phi_{j;m,\bar m}$, with the $(j,m,\bar m)$ indicated in \pzero. Under the reduction \decomphi, this operator does not, in general, reduce to a physical vertex operator in the coset.

Applying the raising operators in \pzero\ corresponds in terms of the vertex operator to multiplying by the appropriate currents. Thus, the vertex operator corresponding to \pzero\ is
\eqn\verosc{(J^+)^l(\bar J^+)^l\Phi_{l-1;-l,-l}~.}
Note that the currents in \verosc\ do not have short distance singularities either among themselves or with the primary $\Phi_{l-1;-l,-l}$. This means that the vertex operator \verosc\ is a primary of Virasoro. Its dimension is given by
\eqn\hn{\Delta(l)=l-{l(l-1)\over k-2}~.}
An interesting question is what is the decomposition of the operator \verosc\ under $SL(2,\IR)/U(1)\times U(1)$. The OPE of the current $J^3(z)$ with \verosc\ can only contain a single pole, from the OPE of $J^3$ with the currents, and with the primary $\Phi_{j;m,\bar m}$. The residue of the pole is the total charge of the operator, which is equal to $l-l=0$. Thus, the operator \verosc\ commutes with the charges $J^3_n$, \ie\ it belongs to the coset $SL(2,\IR)/U(1)$. As we will review below, in the language of \momwin\ it has $p=w=0$.

The dimension of \verosc, \hn, is equal to one that appeared in our discussion before, \dimpurewin, with the mapping $n=l+1$. There, we discussed states with $w=1$, $p=0$, while here we are dealing with states with $w=p=0$. The conserved charge $p$ is the same in the two cases, while $w$ is seemingly different (but is not conserved). We will see below that the agreement of \dimpurewin, \hn\ is not accidental, and  the two are in fact the same due to the generalized FZZ correspondence.

The primary state in \pzero\ $|j=l-1;m=\bar m=-l;w=0\rangle$ is isomorphic, according to our previous discussion, to the state
$|\tilde j={k\over 2}-l-1;m={\bar m}={k\over 2}-l;w=1\rangle$. Acting $l$ times with $J_{-1}^+\bar J_{-1}^+$ corresponds \sflow\ to acting $l$ times with $\tilde J_0^+\bar{\tilde J}_0^+$ on the $w=1$ state, which leads to the state
$|\tilde j={k\over 2}-l-1;m={\bar m}={k\over 2};w=1\rangle$.
The corresponding vertex operators can be constructed from those with $w=0$, \decomphi, by multiplying with certain twist fields (see \eg~\refs{\MaldacenaHW,\ArgurioTB}).
They are given by
\eqn\operators{\Phi_{j;m,\bar m}^w
=V_{j;m,\bar m}e^{\sqrt{2\over k}(m-{k\over 2}w)x(z)+
\sqrt{2\over k}(\bar m-{k\over 2}w)x(\bar z)}~,}
and have dimension and charge \dimjw. For our particular case, $j={k\over2}-l-1$, $m=\bar m={k\over2}$, and the vertex operator \operators\ reduces to one that lives purely in the coset, $V_{{k\over2}-l-1;{k\over2},{k\over2}}$. This operator has winding one \momwin, and according to our discussion of $SL(2,\IR)$ CFT above it creates the same normalizable state as the seemingly different vertex operator \verosc.

In the next section we will study this duality further, and in particular address the question how the two dual vertex operators
\verosc, \operators\ behave at large $\phi$, \ie\ in the region where the cigar can be approximated by a semi-infinite cylinder (see figure 1).

\newsec{Dual vertex operators at large $\phi$}

In the last section we saw that naively different vertex operators \verosc,  \operators, in the Euclidean black hole background \sltwo\ give rise to the same normalizable state in the cigar CFT. The purpose of this section is to provide further insight into this duality by analyzing the asymptotic form of the dual vertex operators far from the tip of the cigar. This will help address some of the questions raised in section 3.

The region far from the tip of the cigar in $SL(2,\IR)/U(1)$ CFT descends from the region near the boundary of $AdS_3$ in the underlying $SL(2,\IR)$ CFT. There is a well known technique for studying this region, known as the Wakimoto representation \WakimotoGF\ (see also \refs{\BernardIY\BershadskyIN-\GiveonNS}). One starts with the worldsheet Lagrangian
\eqn\waklag{\CL=\partial\phi\bar\partial\phi-Q\widehat R\phi+\beta\bar\partial\gamma+\bar\beta\partial\bar\gamma-\lambda\beta\bar\beta e^{-Q\phi}~.}
Integrating out $\beta$ gives the worldsheet $\sigma$-model Lagrangian on $AdS_3$ parametrized by $(\phi,\gamma,\bar\gamma)$. Here $\phi$ is the radial coordinate on $AdS_3$, while $(\gamma,\bar\gamma)$ parametrize the boundary of this spacetime.

The description \waklag\ is particularly useful at large $\phi$, near the boundary of $AdS_3$, where the interaction term goes to zero and the worldsheet theory becomes free. The local string coupling, $g_s(\phi)\sim\exp(-Q\phi/2)$, also goes to zero there, so string interactions are suppressed as well.

In that region, the worldsheet fields $(\phi,\beta,\gamma)$ can be viewed as free fields, with the propagators
\eqn\pbgprop{\langle\phi(z)\phi(0)\rangle=-\ln|z|^2~,\qquad \langle\beta(z)\gamma(0)\rangle={1\over z}~.}
The field $\phi$ corresponding to the radial coordinate of $AdS_3$ behaves as a free field with linear dilaton with slope $-Q/2$. The fields $\beta$ and $\gamma$ are bosonic free fields with dimensions $1$ and $0$ respectively.

The $SL(2,\IR)$ currents are given by
\eqn\jjj{J^+=\beta~,\quad J^3=-\beta\gamma-{1\over Q}\partial\phi~,\quad J^-=\beta\gamma^2+{2\over Q}\gamma\partial\phi+k\partial\gamma~.}
The normalizable primary operators $\Phi_{j;m,\bar m}$ \opejphi\ take at large $\phi$ the form
\eqn\formphi{\Phi_{j;m,\bar m}\simeq \gamma^{-(j+m+1)}\bar\gamma^{-(j+\bar m+1)}e^{-Q(j+1)\phi}~.}
The Wakimoto representation can also be used to determine the form of the vertex operators of $\widehat{SL}(2,\IR)$ descendants. For example, the operators \verosc\ take the form
\eqn\largephides{(J^+)^l(\bar J^+)^l\Phi_{l-1;-l,-l}\simeq (\beta\bar\beta)^le^{-Ql\phi}~.}
Note that for $l=1$ \largephides\ coincides with the interaction term (the last term on the r.h.s.) in \waklag.

Now that we have the large $\phi$ behavior of vertex operators in the $SL(2,\IR)$ CFT, we would like to determine that of their counterparts in the coset model. An efficient technique for doing that was described in \BershadskyIN. It involves adding to the model a $U(1)$ gauge field $(A,\bar A)$, which gives an extra contribution to the $U(1)$ current $J^3$ \defxxx, \jjj, $J^3_A=i\sqrt{k\over2}\partial X$, where $X$ is a canonically normalized scalar field, and there is a similar formula for the right-movers. The total $U(1)$ current,
\eqn\jtotal{J^3_{\rm total}=J^3+J^3_A=i\sqrt{k\over2}\partial(X+ix)~,}
is null. Thus, we can add a pair of fermionic ghosts $b$ and $c$ of dimensions $1$ and $0$ respectively, and construct the BRST charge,
\eqn\nullbrst{Q=\oint cJ^3_{\rm total}~,}
which is nilpotent, $Q^2=0$. The cohomology of the BRST charge \nullbrst\ (and its right-moving analog) is the physical spectrum of the $SL(2,\IR)/U(1)$ coset model. The large $\phi$ form of vertex operators on the cigar can be read off from it.

To see how this works in practice, consider the vertex operator $V_{j;m,\bar m}$, that made an appearance in our discussion above, in the construction of the winding one operator \operators. We can construct this operator by starting with the vertex operator $\Phi_{j;m,\bar m}$, and dressing it with the appropriate gauge field contribution. The dressing is determined by the condition that the total $U(1)$ charge \jtotal\ vanishes, which ensures that the operator is BRST invariant. This leads to the general result
\eqn\dressedphi{V_{j;m,\bar m}=\Phi_{j;m,\bar m} e^{-i\sqrt{2\over k}(mX-\bar m\bar X)}\simeq \gamma^{-(j+m+1)}\bar\gamma^{-(j+\bar m+1)}e^{-Q(j+1)\phi-i\sqrt{2\over k}(mX-\bar m\bar X)}~,}
where in the last equality we used \formphi\ to focus on the large $\phi$ behavior.

As explained in \BershadskyIN, the part of the vertex operator \dressedphi\ that belongs to the $(\beta,\gamma)$ system does not contribute either to the dimension of the operator or to its correlation functions with other BRST invariant operators, and can be omitted. The remaining operator takes the form \behavvv, \plpr.  Thus, we learn that the scalar field $X$ associated with the gauge field is identified with the compact scalar parametrizing the angular direction on the cigar. In particular, it is compact with radius $\sqrt{2k}$, as explained above.\foot{This is related to the fact that the $U(1)$ symmetry that we are gauging is compact.} We also see that the vertex operator $V_{j;m,\bar m}$ describes a tachyon with momentum and winding \momwin.

To study the generalized FZZ correspondence, we need to extend the above discussion to  $\widehat{SL}(2,\IR)$ descendants such as \largephides. We will see that these correspond to oscillator states on the cigar. To demonstrate that, it is convenient to ``bosonize'' the $(\beta,\gamma)$ system in a way familiar from string theory (see \eg\ \PolchinskiRR):
\eqn\bosbg{\beta=-\partial w e^{w-u}~;\qquad \gamma=e^{u-w}~.}
$u$ and $w$ are free fields with $\langle w(z)w(0)\rangle=-\langle u(z)u(0)\rangle=\ln z$. We also have $\beta\gamma=-\partial u$.
Plugging this into \jjj, \jtotal, we find
\eqn\jjttoo{J^3_{\rm total}=\partial u-{1\over Q}\partial\phi+i\sqrt{k\over2}\partial X~.}
Eq. \nullbrst\ implies that
\eqn\jtottriv{J^3_{\rm total}=\{Q,b\}~.}
Thus, in correlation functions of BRST invariant operators we can set the current \jjttoo\ to zero.

We are now ready to discuss the operators \largephides\ as operators in the $SL(2,\IR)/U(1)$ coset, and in particular their form at large $\phi$. Consider first the operator $\beta^l$ written in terms of the bosonized variables $(u,w)$ \bosbg. On general grounds, we know the following:
\item{(1)} Since the OPE of two $\beta$'s does not have a short distance singularity, $\beta^l$ is a Virasoro primary (of dimension $l$).
\item{(2)} In terms of the bosonized variables \bosbg, $\beta^l$ takes the form $P_l(\partial w,\partial^2w,\cdots)e^{l(w-u)}$. $P_l$ is a polynomial in $\partial^n w$, $n=1,2,\cdots$, with total scaling dimension $l$. It can be computed explicitly for all $l$; we will give the result for some low values of $l$ below.
\item{(3)} The operator \largephides\ thus takes the form
\eqn\oscstate{\beta^le^{-lQ\phi}=P_l(\partial w,\cdots) e^{l(w-u-Q\phi)}~,}
where we again suppressed the right-moving part of the operator, which is very similar.
\item{(4)} Both the polynomial $P_l$ and the exponential in \oscstate\ commute with $J^3_{\rm total}$ \jjttoo. Since the latter is BRST exact \jtottriv, in correlation functions of BRST invariant operators we can replace $\partial w\to \partial w - J^3_{\rm total}=\partial w-\partial u+{1\over Q}\partial\phi-i\sqrt{k\over2}\partial X$. The first two terms depend on the combination $w-u$ and thus do not contribute to correlation functions for the same reason as the exponentials of $w-u$ in \dressedphi, \oscstate. Thus, in the polynomial $P_l$ in \oscstate\ we can replace $\partial w$ by the combination ${1\over Q}\partial\phi-i\sqrt{k\over2}\partial X$.

\noindent
To summarize, we conclude that the vertex operator \largephides\ takes in the cigar CFT the large $\phi$ form
\eqn\largephiosc{(\beta\bar\beta)^le^{-Ql\phi}\simeq P_l(\partial w,\cdots)P_l(\bar\partial\bar w,\cdots)e^{-Ql\phi}~,}
with $\partial w\to{1\over Q}\partial\phi-i\sqrt{k\over2}\partial X$. Note that at large $k$, the combination $\partial w$ takes the form
\eqn\largekform{\partial w\to {1\over Q}\partial\phi-i\sqrt{k\over2}\partial X\simeq\sqrt{k\over2}\partial(\phi-iX)=i\sqrt{k\over2}\partial Z~,}
where in the last equality we defined the complex coordinate on the asymptotic cylinder, $Z=\phi-iX$.

The polynomials $P_l$ are in general non-trivial. For $l=1,2,3$ one finds\foot{We neglect an overall $l$-dependent sign which cancels between the two $P_l$'s in \largephiosc.}
\eqn\formpl{P_1=\partial w~,\qquad
P_2=(\partial w)^2-\partial^2w~,\qquad
P_3=(\partial w)^3-3\partial^2w\partial w+\partial^3w~.}
For general $l$, $P_l$ is given by a linear combination of many different terms ranging from $(\partial w)^l$ to $\partial^l w$, with $k$-independent coefficients. In the large $k$ limit, $\partial w$ is naturally of order $\sqrt k$ (see \largekform). Thus, in this limit the polynomial $P_l$ simplifies,
\eqn\simplargek{P_l\simeq(\partial w)^l\simeq k^{l\over2}(\partial Z)^l~.}
The vertex operator \largephiosc\ describes an oscillator state at level $l$. Its GFZZ dual has winding one and $j={k\over 2}-l-1$. The corresponding vertex operator, \dressedphi, has the form
\eqn\wonever{e^{-Q({k\over2}-l)\phi-i\sqrt{k\over 2}(X-\bar X)}~.}
It describes the lowest lying state of a string with winding one around the circle (the winding tachyon).

The two dual operators \largephiosc\ and \wonever\ correspond naively to different modes of the string, and have quite different behaviors at large $\phi$. Nevertheless, they describe the same normalizable state in the theory.  Consider, for example, the special case $l=1$.
In that case the winding tachyon vertex operator \wonever\ reduces to the Sine-Liouville operator \refs{\fzz,\KazakovPM}. The oscillator vertex operator \largephiosc\ becomes the metric deformation that closes up the infinite cylinder to a cigar (the leading expansion of the metric \sltwo\ at large $r$). The identification between the two is the original FZZ correspondence.

One can view the arguments presented above in two different ways. If one takes the identification of the $\widehat{SL}(2,\IR)$ representations described in section 4 as given, these arguments provide a derivation of the (G)FZZ correspondence from properties of the underlying $SL(2,\IR)$ WZW model. Conversely, if one takes the FZZ correspondence as given, they provide strong evidence for the identification of representations with $w=0$ and $w=1$ described in the previous section, at least for the special case $l=1$. This makes it natural to identify the $\widehat{SL}(2,\IR)$ representations with $l>1$ as well, which leads to the GFZZ correspondence.

\newsec{Comments on the correspondence}

In sections 3 -- 5 we discussed a class of normalizable states on the cigar. We saw that these states have two components, which  from the perspective of the asymptotic cylinder geometry have windings zero and one, respectively. The two components are always present; in general they have different localization properties in the radial direction and so influence physics at different scales. We referred to this as the generalized FZZ correspondence.

In section 3 we discussed the winding one component of these states. The fact that we did not include the winding zero component led to an incomplete picture. In this section we would like to describe these states taking into account the GFZZ correspondence, which will give a more complete picture.

We start by considering low lying states, corresponding to $l\sim O(1)$, while taking $k$ to be large.  The $w=0$ contribution to such a state is given by the vertex operator \largephiosc, \simplargek,
\eqn\wone{ (\partial Z)^l (\bar{\partial}\bar{Z})^l e^{-Ql\phi}~.}
This vertex operator describes an oscillator state whose zero mode wavefunction is supported in a region (roughly) the size of the curvature radius of the cigar $(\sqrt k)$ (see figure 4).

The $w=1$ contribution \wonever\ to the same state is highly localized near the tip of the cigar (figure 4), where the potential experienced by the wound string is quadratic in the radial coordinate. Thus, it gives rise to a two dimensional harmonic
oscillator~\refs{\GiveonICA,\GiveonHFA}.
The states that correspond to \wone\ take the form
\eqn\zero{ (a_+^{\dag})^l (a_-^{\dag})^l |0,0\rangle~,}
where $a_+^{\dag}$ and $a_-^{\dag}$ are the two creation operators associated with the harmonic oscillator.

\ifig\loc{The two components of each normalizable state have in general different localization properties. For low lying states, the oscillator state contribution (a) is extended over a larger region than the winding tachyon one (b). }
{\epsfxsize5in\epsfbox{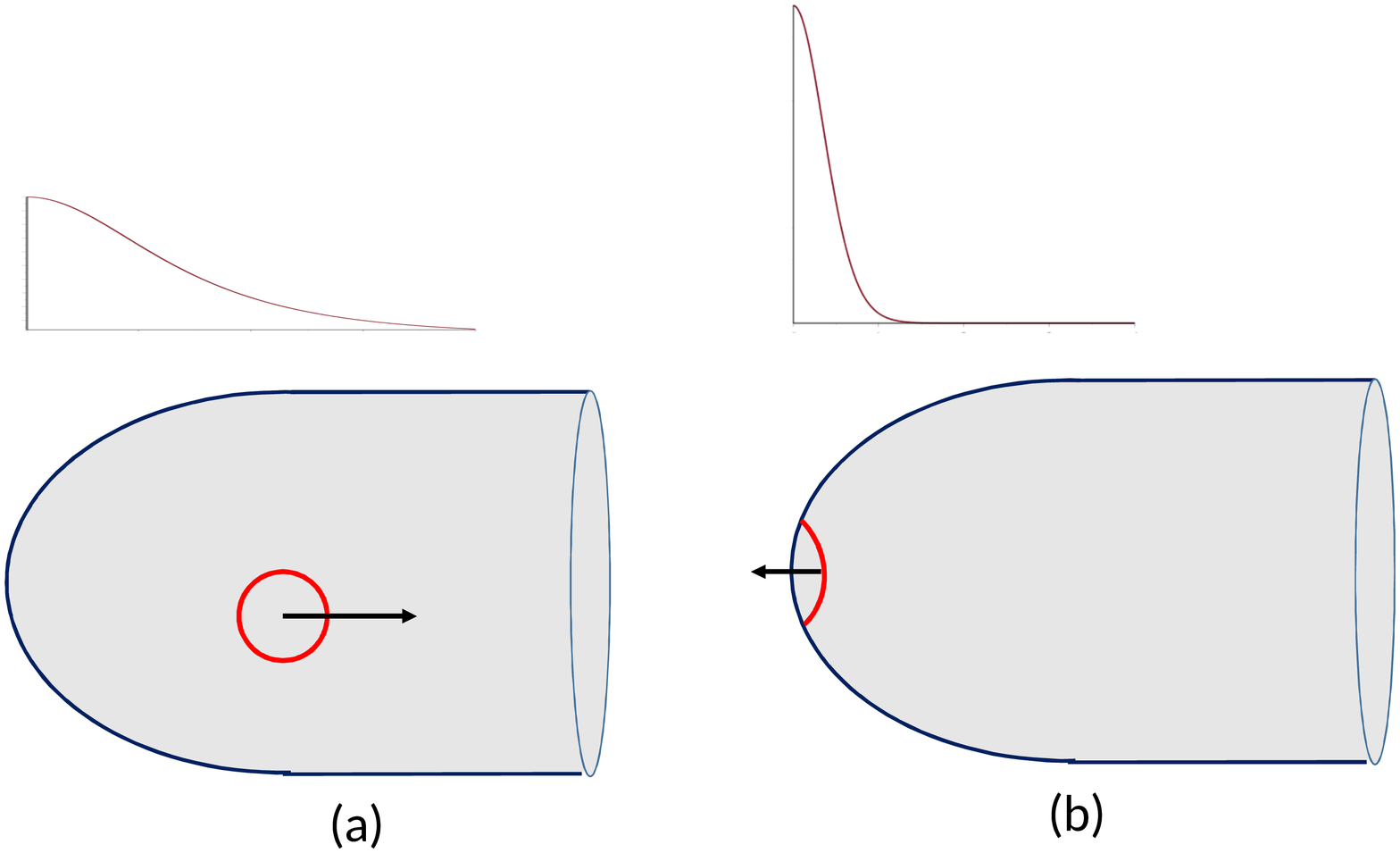}}

The structure of \zero\ is very similar to that of \wone, with the role of the worldsheet oscillators $(\alpha_{-1},\bar\alpha_{-1})$ of $(Z,\bar Z)$ in \wone\ played by the spacetime oscillators $(a_+^{\dag},a_-^{\dag})$. For a string wrapping the circle on the  cigar the two are closely related, as can be seen by choosing the gauge $\theta(\sigma)=\sigma$.

We see that the states \wone, which correspond to standard oscillator states of a string whose wavefunction is spread over the cap of the cigar (see figure 1), have a  component \zero\ that has the same oscillator structure, but is localized at the tip. In the next section we shall see that this picture remains intact when we turn on angular momentum. Low energy probes are insensitive to the localized contribution and experience only the oscillator component \wone. For $l=1$ it was shown in  \GiveonCMA\ that energetic probes are quite sensitive to  \zero. We expect this to remain the case for $l>1$ as well.

Note that in \zero, like in \wone, the parameter $l$ takes value in the positive integers. In particular, the ground state of the harmonic oscillator, which corresponds to $l=0$, does not give rise to a physical state. The reason for that is mysterious from the perspective of section 3, but is easy to understand from the GFZZ correspondence -- the dual state \wone\ with $l=0$ is not normalizable. This is an example of a fact mentioned in section 3: at small $l$, one expects the winding string description of that section to be subject to significant corrections.

As $l$  increases, the $w=0$ contribution to the state  undergoes two processes. The zero mode wavefunction becomes more localized -- it is supported in the region $\phi\le{1\over Ql}$ -- while the increasing oscillator level leads to an increase in the size of the string, which goes like $\sqrt l$, due to the fact that a string at oscillator level $l$ can be viewed as a random walk with $l$ steps. Eventually, when $l$ becomes of order $k$ this picture breaks down, since the size of the string becomes comparable to the radius of curvature of the geometry.

The $w=1$ contribution to the state also undergoes two processes as $l$ increases. The zero mode wavefunction in \wonever\ spreads to larger $\phi$, $\phi\le {1\over Q({k\over 2}-l)}$. The size of the string again grows, initially like $\sqrt l$, which in this language is due to the growth of harmonic oscillator states with the level. As $l$ continues to increase, one eventually reaches a regime in which the wound string probes the region in its potential where the harmonic approximation breaks down, and one is sensitive to the full potential described in section 3.  The flattening of the potential at infinity gives rise to an upper bound on $l$, and modifies the zero mode wavefunctions from their harmonic oscillator forms.

As $l$ approaches the top of its range, $l\sim k/2$, the size of the winding string state approaches $\sim\sqrt k$. In this region, the semiclassical description of section 3 (and its exact analog in section 5, given by the vertex operator \wonever) becomes accurate.

The $w=0$ contribution to the state \wone\ has the following behavior for $l\sim k/2$. The center of mass wavefunction is sharply peaked at small $\phi$ -- at large $\phi$ it goes to zero rapidly, like $\exp(-\phi/Q)$. However, the oscillator level $l$ in \largephiosc\ is large in this case, and the size of the string behaves like $\sqrt{k}$, due to the random walk.

It is interesting that the size of the oscillator state predicted by the random walk picture is comparable to the size of the winding string with the same value of $l$ predicted by the analysis of sections $3-5$. We would like to argue that this is not a coincidence.  The oscillator state \wone\ ``knows about'' the dual winding string state \wonever. The long random walk with $l$ steps provides a description of the winding string from the perspective of the oscillator state. However, that description has large fluctuations, associated with the random walk. The winding description is better in this regime ($l\sim k/2$) in the sense that it is semiclassical -- the string that winds around the cigar is straight, and does not have large fluctuations. This property is not easy to see from the oscillator state perspective.

One way to summarize the above discussion is by describing the  properties of the $l$'th normalizable state as viewed by a low energy observer as a function of $l$. For small $l$ the good description of this state is as an oscillator state \wone. Its size is governed by the zero mode wavefunction; it goes like $\sqrt{k}/l$, thus decreasing with $l$. At intermediate values of $l$ the situation is in general complicated, as one needs to take into account both the winding and momentum components of the normalizable state, and the growth of the size of the oscillator state with $l$. When $l$ approaches $k/2$, \ie\ when $\alpha$ in \largenn\ is of order one, the correct long distance description is the winding one, and the size of the state grows with $l$ in the way described after eq. \decayv.

The above discussion is reminiscent of that of the string/black hole transition in \refs{\HorowitzNW,\HorowitzJC,\KutasovRR,\GiveonPR}, who discussed the question what happens to an excited string state as we increase its excitation level, eventually reaching masses for which the corresponding black hole is large. The picture proposed in these papers is that when the excitation level is such that the corresponding black hole has a horizon of size $l_s$, the description in terms of perturbative string states with their Hagedorn entropy is replaced by that in terms of classical black holes and Bekenstein/Hawking entropy. The mechanism driving this transition is quantum $(g_s$) effects, which make the naively large highly excited perturbative string state shrink as we approach the correspondence point, beyond which the states start growing again according to the black hole description.

In our case, we have a similar transition that happens as a function of the oscillator level $l$, and like in the case of the string/black hole transition, as we increase $l$ the states first decrease and then increase in size. Instead of large black holes, here we have long winding strings, and instead of $g_s$ effects that drive the transition there, here it is (non-perturbative) $\alpha'$ effects associated with the generalized FZZ correspondence. The reason that we can say more about the transition in our case than is currently possible in the string/black hole case is that unlike the $g_s$ effects there, the $\alpha'$ effects are under complete control here.

\newsec{Generalizations}

In this section we discuss some generalizations of the basic idea presented in the previous sections.

\subsec{Generic $p$ ($w=0$ vs. $w=1$)}

In the previous sections we saw that normalizable states on the cigar with zero momentum around the cigar have two components, which from the perspective of the asymptotic cylinder have winding zero and one, respectively. In this subsection we show that this is the case for non-zero momentum around the cigar as well.

A large set of states in the representation $\hat D_j^{-,w=0}\otimes \bar{\hat D}_j^{-,w=0}$, which reduce to primaries with $w=0$ and generic momentum $p$ around the cigar, are given by
\eqn\pnonzero{(J_{-1}^+)^l(\bar J_{-1}^+)^{\bar l}
|j={1\over 2}(l+\bar l)-1;m=\bar m=-(j+1);w=0\rangle~.}
They correspond to $\widehat{SL}(2,\IR)_L\times \widehat{SL}(2,\IR)_R$ currents acting on operators $\Phi_{j;m,\bar m}^w$
(with the corresponding $(j;m,\bar m;w)$);  in the Wakimoto variables they take the form
\eqn\wakpnonzero{(J^+)^l(\bar J^+)^{\bar l}\Phi^{w=0}_{j={1\over 2}(l+\bar l)-1;m=\bar m=-(j+1)}\sim
\beta^l\bar\beta^{\bar l}e^{-{1\over 2}Q(l+\bar l)\phi}~.}
These reduce on the asymptotic cylinder in figure 1 to
\eqn\plpbarl{P_l(\partial w,\cdots)P_{\bar l}(\bar\partial\bar w,\cdots)
e^{i{p\over\sqrt{2k}}X}e^{-{1\over 2}Q(l+\bar l)\phi}~,}
namely, to
\eqn\cigarpnonzero{((\partial Z)^l+\dots)((\bar\partial\bar Z)^{\bar l}+\dots)
e^{i{p\over\sqrt{2k}}X}e^{-{Q\over 2}(l+\bar l)\phi}~,}
where
\eqn\ppp{p=\bar l-l}
is the angular momentum on the cigar, and the ``$\dots$'' in \cigarpnonzero\ stand for $1/k$ corrections, which in particular make \cigarpnonzero\ a primary. In the special case $p=0$ the above states reduce to those described in sections 4,5. In particular, \pnonzero\ generalizes \pzero, \wakpnonzero\ generalizes \largephides, and \plpbarl\ -- \ppp\ generalize \largephiosc, \simplargek.

The states \pnonzero\ are isomorphic to certain states in the
$\hat D^{+,w=1}_{\tilde j={k\over 2}-j-2}\otimes
\bar{\hat D}^{+,w=1}_{\tilde j={k\over 2}-j-2}$ representation with
\eqn\jjj{j+1={1\over 2}(l+\bar l)~,}
concretely, to
\eqn\wonepzero{|\tilde j={k\over 2}-j-2~;
(m,\bar m)={1\over 2}(k-p,k+p);w=1\rangle~,}
which correspond to the operators \operators\
with the corresponding $(\tilde j;m,\bar m;w)$.
On the asymptotic cylinder of the cigar these reduce to
\eqn\cigarwone{e^{-Q(\tilde j+1)\phi}
e^{-i\sqrt{2\over k}(m X(z)-{\bar m}X(\bar z))}~,}
where
\eqn\tildejmm{\tilde j+1={k-l-\bar l\over 2}~,\qquad
(m,{\bar m})={1\over 2}(k-p,k+p)~.}
They have $w=1$ and momentum $p$ around the cigar.

At small $\phi$, near the tip of the cigar, they are described by the following states in the 2d harmonic oscillator
\eqn\zer{ (a_+^{\dag})^l (a_-^{\dag})^{\bar{l}} |0,0\rangle~.}
Comparing \zer\ with \cigarpnonzero\ we see that, just like in the $p=0$ case, in the large $k$ limit the oscillator structures of the diffuse and localized components agree.

\subsec{Generic $(p,w)$}

The isomorphism between $\widehat{SL}(2,\IR)_L\times \widehat{SL}(2,\IR)_R$ representations
reviewed in section 4 is a particular case of a more general isomorphism \MaldacenaHW,
relating
\eqn\djdj{\hat D_j^{-,w}\otimes {\hat D}_j^{-,w}\longleftrightarrow
\hat D_{{k\over 2}-j-2}^{+,w+1}\otimes {\hat D}_{{k\over 2}-j-2}^{+,w+1}~.}
This gives rise, upon reduction to the $SL(2,\IR)/U(1)$ CFT, to a GFZZ duality between operators corresponding to states with generic winding and momentum. For $w>0$,
the left and right dimensions of such states, $\Delta$ and $\bar\Delta$, are of order $k$. Hence, the only case for which the GFZZ correspondence affects physics at energies well below $k$ is the duality between $w=0,1$ described above. The dualities \djdj\ with $w\ge 1$ play a role at high energies, and for $k$ of order $1$.

\subsec{$\CN=2$ superconformal case}

As mentioned in section 2, the supersymmetric extension of the bosonic cigar theory gives rise to an $\CN=2$  SCFT. The $\CN=2$ superconformal symmetry is very useful for organizing the spectrum of the theory into $\CN=2$ primaries and descendants. The latter dominate the high energy density of states.

The $\CN=2$ superconformal symmetry also provides additional evidence for the GFZZ correspondence \GIKthree.
All the discrete representations of the underlying $SL(2,\IR)$ SCFT contain a state
which reduces to a BPS state,
whose $\CN=2$ character contributes to the elliptic genus of the
supersymmetric cigar theory.
This elliptic genus has been studied in the literature for integer $k$
(see \eg\ \refs{\SugawaraHMA,\GiveonHSA,\GiveonHFA})
and is known including overall normalization (which can be obtained \eg\ from the Witten index). One can then ask whether the BPS states contribute with coefficient one to the elliptic genus, or two, for the state and its GFZZ dual. One finds that the answer is one, which implies that the two states are indeed identified.

Once these states are identified in the superconformal $SL(2,\IR)/U(1)$ theory, this must also be true for their ancestors in the underlying $\CN=1$ superconformal $SL(2,\IR)$  CFT. The latter consists of a bosonic $SL(2,\IR)$  CFT and three free fermions in the adjoint of $SL(2,\IR)$. Thus, the identification of representations necessary for the elliptic genus to work also implies the identification necessary for the bosonic coset model.

\newsec{Summary and Discussion}

In this note we showed that standard normalizable states in string theory on the cigar \sltwo\ that are smeared over the cap (the region with curvature $\sim 1/k$) in figure 1, have a component that is localized a stringy distance from the tip. Despite the fact that their target-space description is so different, in the CFT  the two components cannot be separated. This suggests that the information that from one perspective is smeared over the whole cap, from the other is localized near the tip. Reference \refs{\GiveonCMA} suggests that low-energy probes are sensitive only to the standard stringy modes while high-energy modes are sensitive also to the modes that are localized at the tip.

It is natural to wonder what might be the implications of our results for Lorentzian black holes. The Wick rotation takes the cap of the cigar to the black hole atmosphere (the region outside the horizon where the potential is attractive towards the horizon), and the tip to the black hole horizon. This suggests that, at least naively, in string theory  the  information in the black hole atmosphere is stored also %xeroxed
at the black hole horizon (and possibly in the interior, that is absent in the Euclidean geometry). To make this exciting possibility precise one needs to understand the analog of the generalized FZZ correspondence for the  Lorentzian black hole.

\bigskip
\noindent{\bf Acknowledgements:}
We thank R. Wald for discussions.
The work of AG and NI is supported in part by the I-CORE Program of the Planning and Budgeting Committee and the Israel Science Foundation (Center No. 1937/12), and by a center of excellence supported by the Israel Science Foundation (grant number 1989/14). DK is supported in part by DOE grant DE FG02-13ER41958. DK thanks Tel Aviv University and the Hebrew University for hospitality during part of this work.

\listrefs

\end